\documentclass[11pt]{article}
\usepackage[english]{babel}
\usepackage{amsfonts,amsmath,amsxtra,amsthm,amssymb,latexsym}
\usepackage[cp866]{inputenc}
\usepackage{graphicx}

\renewcommand{\Re}{\mbox{Re}}

\newcommand{\We}{\mbox{We}}
\newcommand{\R}{\mathbb{R}}

\makeatletter\@addtoreset {equation}{section}\makeatother
\theoremstyle{plain}

\begin{document}

\title{Asymptotic behavior of regularized shock solutions in coating flows}

\author{D. Badali$^1$, M. Chugunova$^2$, D.E. Pelinovsky$^3$, and S. Pollack$^4$ \\
{\small $^{1}$ Department of Chemical and Physical Science,
University of Toronto at Mississauga,}\\{\small Mississauga, Ontario, Canada, L5L 1C6}\\
{\small $^{2}$ Department of Mathematics, University of Toronto, Toronto, Ontario, Canada, M5S 1A1}\\
{\small $^{3}$ Department of Mathematics, McMaster University, Hamilton, Ontario, Canada, L8S 4K1}\\
{\small $^{4}$ Department of Mathematics, McGill University, Montreal, Quebec, Canada, H3A 2T5} }

\date{}
\maketitle

\begin{abstract}
We consider a model for thin liquid films in a rotating cylinder
in the small surface tension limit. Using dynamical system methods, we show
that the continuum of increasing shock solutions persists in the small surface tension limit,
whereas the continuum of decreasing shock solutions terminates at the limit.
Using delicate numerical computations, we show that the existence curves of
regularized shock solutions on the mass-flux diagram exhibit loops.
The number of loops increases and their locations move to infinity as
the surface tension parameter decreases to zero.
If $n$ is the number of loops in the mass-flux diagram with $2n+1$ solution branches,
we show that $n+1$ solution branches are stable with respect to small perturbations.
\end{abstract}

\section{Introduction}

The time evolution of a liquid film spreading over
a solid surface under the action of the surface tension and
viscosity can be described by lubrication models \cite{AHS2003,ESR2004,John,SW1995}.
These models approximate the full
Navier-Stokes equations that appear in the study of motion and
instabilities of the liquid film dynamics. Thin films
play an increasingly important role in a wide range of applications,
for example, packaging, barriers, membranes, sensors, semiconductor
devices, and medical implants. Spin coating is one of the methods
that is widely used to coat uniform thin films onto solid surfaces
in a variety of industrial applications such as manufacturing of the magnetic
and optical discs. In this technology, a liquid drop spreads radially
due to centrifugal effects from spinning and eventually yields a
thin film of uniform thickness formed on the solid surface. In
experiments, a variety of different types of bifurcations and steady
states have been observed \cite{DW1999,HK1994,TM1997}.

In this paper we consider the dynamics of a viscous incompressible
thin fluid film on the outer surface of a horizontal circular
cylinder that is rotating around its axis in the presence of
a gravitational field. The coating flow is generated by
viscous forces due to the cylinder's surface motion relative to the
fluid. There is no temperature gradient, hence the interface does
not experience a shear stress. If the cylinder is fully coated there
is only one free boundary: where the liquid meets the surrounding
air. Otherwise, there is also a free boundary (or contact line)
where the air and liquid meet the cylinder's surface. The motion of
the liquid film is governed by four physical effects: viscosity,
gravity, surface tension, and centrifugal forces. These are
reflected in the following parameters:
\begin{itemize}
\item $R$ is the radius of the cylinder;
\item $\omega$ is the constant rate of its rotation;
\item $g$ is the acceleration due to gravity;
\item $\nu$ is the kinematic viscosity;
\item $\rho$ is the fluid density;
\item $\sigma$ is the surface tension.
\end{itemize}
These parameters yield three independent dimensionless numbers:
\begin{equation}
{\rm Re} = \frac{R^2 \omega}{\nu}, \quad \gamma = \frac{g}{R \omega^2}, \quad
{\rm We} = \frac{\rho R^3 \omega^2}{\sigma},
\end{equation}
where ${\rm Re}$ is the Reynolds number and ${\rm We}$ is the Weber number.

Taking the ratio $\epsilon = \bar{h}/R$ as a small parameter, where
$\bar{h}$ is the average thickness of the liquid, we consider the limit $\epsilon \to 0$ such that
\begin{equation}
\label{params}
\chi =  \tfrac{\Re}{\We} \,\epsilon^3 \quad \mbox{and} \quad  \mu = \gamma \, \Re \,
\epsilon^2
\end{equation}
remain finite and nonzero.

One can model the flow on a rotating cylinder using the full
Navier-Stokes equations for the velocity vector $\vec{u}(r,\theta,z,t)$,
where $r \in [0,h(\theta,t)]$ is the axial variable,
$\theta \in [-\pi,\pi]$ is the angular variable, $z \in \R$ is the
variable in the direction of the cylinder, and $h(\theta,t)$ is the thickness of the fluid
on the surface of the cylinder at time $t$. The simplifying model
that takes into account the small surface tension and the gravitational force
was considered in a number of works including Pukhnachev \cite{Pukh1} and O'Brien \cite{OBrien}.
This model is written in the form,
\begin{equation} \label{A:PukhEq}
\partial_t h + \partial_{\theta} \left[ h - \tfrac{1}{3} \mu h^3 \, \cos(\theta) \right] +
\tfrac{1}{3} \chi \partial_{\theta}
\left[ h^3 \left( \partial_{\theta}  h + \partial_{\theta}^3  h \right) \right] = 0,
\end{equation}
where $\mu$ and $\chi$ are given in (\ref{params}) and  $h(\theta + 2\pi,t) = h(\theta,t)$. The
model assumes no-slip boundary conditions at the liquid/solid
interface. A solution to equation (\ref{A:PukhEq})
is physically relevant if either $h$ is strictly positive (the
cylinder is fully coated) or $h$ is nonnegative (the cylinder is wet
in some region and dry in others).

In a similar context of the syrup rings on a rotating
roller, Moffatt \cite{Moff} neglected the effects of the surface tension (i.e.,
$\We^{-1}=0 = \chi$) and obtained the reduced equation,
\begin{equation}
\label{A:MoffEq} \partial_t h + \partial_{\theta} \left[ h - \tfrac{1}{3} \mu h^3 \, \cos(\theta)
\right] = 0.
\end{equation}

The stationary solutions of (\ref{A:PukhEq}) are given by the $2\pi$-periodic
solutions of the third-order differential equation,
\begin{equation}
\label{stat-eq}
h - \tfrac{1}{3} \mu h^3 \, \cos(\theta) +
\tfrac{1}{3} \chi h^3 \left( \partial_{\theta} h  + \partial_{\theta}^3 h \right) = Q,
\end{equation}
where $Q$ is the constant that corresponds physically to flux of the liquid
through the film cross section.

Moffatt \cite{Moff} and O'Brien \& Gath \cite{O'BG1998} considered stationary solutions
of (\ref{stat-eq}) with $\chi = 0$. Besides smooth periodic solutions for small values of $Q$,
there are two continua of shock solutions for a critical value of $Q = Q_c$, one is associated
with the increasing shocks and the other one is associated with the decreasing shocks. Both families of
solutions are parameterized by the integral
\begin{equation}
\label{mass}
M = \frac{1}{2\pi} \int_{-\pi}^{\pi} h(\theta) d \theta,
\end{equation}
which has the physical meaning of the mass of the liquid. The solution branches can be plotted on the
parameter plane $(M,Q)$ for fixed values of $\mu$ and $\chi$, which we term as the {\em mass--flux diagram}.

Physical arguments convinced the authors of \cite{O'BG1998} that the decreasing shocks are unstable
and cannot be observed in the rotational cylinder if the surface tension effects are included
with $\chi \neq 0$. The singular perturbation theory of small $\chi$ was recently considered
by Benilov {\em et al.} \cite{Benilov2}, where the authors showed that the decreasing shocks do not exist for
small positive $\chi$. On the other hand, increasing shocks become regularized for small $\chi > 0$ and
asymptotic arguments complemented by the numerical approximations were developed in \cite{Benilov2}
to predict spectral stability of stationary solutions with regularized increasing shocks.

Pukhnachev \cite{Pukh3} proved the existence and uniqueness of the steady states in the
differential equation (\ref{stat-eq}) if $\chi$ and $Q$ are not too large. Karabut \cite{Kar}
constructed two branches of steady states in the opposite limits of large $\chi$. Numerical approximations
in Benilov {\em et al.} \cite{Benilov3} (Figure 14) showed that the mass-flux diagram may become
more complicated for small $\chi$ and large $M$ and may include a loop near the value
$Q = Q_c$. Three solutions coexist for a fixed $M$ if the loop is present.
Stability of these solutions was not studied in \cite{Benilov3}.

It is the purpose of this work to continue, improve, and clarify the preliminary results of
Benilov {\em et al.} \cite{Benilov2,Benilov3}. In particular, we develop the dynamical system methods to prove
that the family of increasing shocks persists with respect to $\chi \neq 0$, whereas the family
of decreasing shocks terminate at $\chi = 0$. We develop a delicate numerical approximation
of the steady solutions of equation (\ref{stat-eq}) to show that the number of loops on the mass-flux
diagram increases when $\chi$ is reduced to zero and the location of these loops go to infinity.
We also apply numerical approximations of eigenvalues of the linearized time evolution associated
with the lubrication model (\ref{A:PukhEq}) and show that if $n$ is the number of loops on the
mass-flux diagram, then $n+1$ solution branches are stable with respect to small perturbations.

The article is organized as follows. Section 2 presents results
of the geometric theory on persistence of increasing regularized shocks.
Section 3 presents numerical results on multi-valued loops in the mass--flux diagram.
Section 4 provides a summary and discusses open questions.

{\bf Acknowledgments}: This work was carried out during the Fields-MITACS Undergraduate Summer Research Program
in 2010. The authors thank A. Kulyk for collaborations during this program.

\section{Geometric theory of regularized shocks}

We shall study here asymptotic solutions of the steady-state equation (\ref{stat-eq}).
Using the transformation $h = \tilde{h} \mu^{-1/2}$, $Q = \tilde{Q} \mu^{-1/2}$, and
$\chi = \epsilon \mu^{3/2}$ and dropping the tilde sign for $h$ and $Q$, we obtain
the third-order differential equation
\begin{equation}
\label{third-order}
\epsilon \left[ \frac{d^3 h}{d \theta^3} + \frac{d h}{d \theta} \right] =
\cos(\theta) - \frac{3(h - Q)}{h^3}, \quad \theta \in (-\pi,\pi),
\end{equation}
where $\epsilon > 0$ and $Q > 0$ are parameters of the problem.

Solutions of the limiting problem
\begin{equation}
\label{algebraic-order}
F_Q(h) := \frac{3(h - Q)}{h^3} = \cos(\theta), \quad \theta \in (-\pi,\pi),
\end{equation}
depends on the value of the flux $Q > 0$. If $Q \in \left(0,\frac{2}{3}\right)$,
the $2\pi$-periodic solution $h(\theta)$ is unique \cite{Moff}.

Let us denote the smallest roots of $F_Q(h) = \pm 1$ for $Q \in \left(0,\frac{2}{3}\right)$
by $h_{\pm}$ such that $h_- < h_+$. The unique solution of the limiting problem (\ref{algebraic-order})
satisfies
\begin{equation}
\label{properties-h}
h(-\theta) = h(\theta) : \quad h'(\theta) > 0, \quad \theta \in (-\pi,0),
\end{equation}
with $h(\pm \pi) = h_-$, $h(0) = h_+$, and $h'(\pm \pi) = h'(0) = 0$.

For $Q > \frac{2}{3}$, no solution $h(\theta)$ exists because
$\max\limits_{h \in \mathbb{R}_+} F_Q(h) < 1$.

For $Q = Q_* = \frac{2}{3}$, there is a unique continuous solution
$h(\theta)$ with properties (\ref{properties-h}) and $h(\pm \pi) = h_-$, $h(0) = h_+ \equiv 1$,
$h'(-\pi) = 0$, and $\lim\limits_{\theta \to -0} h'(\theta) = \frac{1}{\sqrt{6}}$.

Besides this continuous solution at $Q = Q_* = \frac{2}{3}$, there exists two symmetric families of shock
solutions with a jump discontinuity at either $\theta = \theta_0$ or $\theta = -\theta_0$, where $\theta_0 \in \left(0,\frac{\pi}{2}\right)$ is a continuous parameter.

Let us denote the two simple zeros of $F_{Q_*}(h) = \cos(\theta_0) \in (0,1)$ by $H_{\pm}$
such that $H_- < H_+$. The increasing shock is centered at $\theta = -\theta_0$ and satisfies
\begin{equation}
\label{properties-shock}
h'(\theta) > 0, \quad \theta \in (-\pi,-\theta_0), \quad h'(\theta) < 0, \quad \theta \in (-\theta_0,\pi),
\end{equation}
with $h(\pm \pi) = h_-$, $\lim\limits_{\theta \to \theta_0-0} h(\theta) = H_-$, $\lim\limits_{\theta \to \theta_0+0} h(\theta) = H_+$, and $h'(\pm \pi) = 0$.

Using the symmetry the limiting problem (\ref{algebraic-order}) with respect to reflection $\theta \to -\theta$,
the decreasing shock can be constructed using the reflection. It is then centered at $\theta = \theta_0$.

The net mass $M$ defined by (\ref{mass}) is a one-to-one increasing function of $Q$ for $Q \in \left(0,Q_*\right)$
with $\lim\limits_{Q \to 0} M = 0$ and $\lim\limits_{Q \to Q_* - 0} M = M_*$ for some $M_* < \infty$,
whereas the two families of shock solutions for $Q = Q_*$ correspond to the values of
$M \in (M_*,\infty)$.

To consider the persistence of the two shock solutions with respect to parameter $\epsilon$, we shall zoom the
coordinate $\theta$ near $\pm \theta_0$ by the transformation
\begin{equation}
\label{scaling-transf}
h(\theta) = H(x), \quad x = \frac{\theta \mp \theta_0}{\epsilon^{1/3}}.
\end{equation}
The new function $H(x)$ satisfies a new version of the third-order differential equation
\begin{equation}
\label{third-order-rescaled}
\frac{d^3 H}{d x^3} + \epsilon^{2/3} \frac{d H}{d x} = \cos(\pm \theta_0 + \epsilon^{1/3} x) - F(H), \quad
x \in \left( \frac{-\pi\mp \theta_0}{\epsilon^{1/3}},\frac{\pi\mp \theta_0}{\epsilon^{1/3}} \right),
\end{equation}
where $H(x)$ is a periodic function with period $2\pi \epsilon^{-1/3}$ and
$$
F(H) = F_{Q_*}(H)=\frac{3H - 2}{H^3}.
$$

The limiting problem at $\epsilon = 0$ becomes now the autonomous equation,
\begin{equation}
\label{third-order-limit}
\frac{d^3 H}{d x^3} = \cos(\theta_0) - F(H), \quad
x \in \mathbb{R}.
\end{equation}

Recall that $H_{\pm}$ are zeros of $F(H) = \cos(\theta_0)$ with ordering $H_- < H_+$.
The increasing shock $h(\theta)$ corresponds
to a heteroclinic orbit $H(x)$ of the limiting problem (\ref{third-order-limit})
satisfying the boundary conditions
\begin{equation}
\label{increasing}
\lim_{x \to \pm \infty} H(x) = H_{\pm}.
\end{equation}
The decreasing shock corresponds to a heteroclinic orbit with the
boundary conditions
\begin{equation}
\label{decreasing}
\lim_{x \to \pm \infty} H(x) = H_{\mp}.
\end{equation}

Linearization of the limiting equation (\ref{third-order-limit}) near the equilibrium states
$H_{\pm}$ gives
\begin{equation}
\label{linearized}
\frac{d^3 \tilde{H}_{\pm}}{d x^3} = - F'(H_{\pm}) \tilde{H}_{\pm},
\end{equation}
where $F'(H_-) > 0$ and $F'(H_+) < 0$. Therefore, the equilibrium state
$H_-$ has a {\em two-dimensional} unstable manifold $W^u(H_-)$ and a {\em one-dimensional}
stable manifold $W^s(H_-)$, whereas the equilibrium state
$H_+$ has a {\em one-dimensional} unstable manifold $W^u(H_+)$ and a {\em two-dimensional}
stable manifold $W^s(H_+)$.

Intersection of the two-dimensional manifolds in $W^u(H_-) \cap W^s(H_+)$ is transverse in the space $\mathbb{R}^3$. Hence, a homoclinic
 orbit satisfying the boundary conditions (\ref{increasing}) exists generally and
persists under the perturbation.

Intersection of the one-dimensional manifolds in $W^u(H_+) \cap W^s(H_-)$ is non-transverse in the space $\mathbb{R}^3$. Hence, a homoclinic orbit satisfying the boundary conditions (\ref{decreasing}) does not exist generally and does not persist under the perturbation. Moreover, it was shown in \cite{Benilov2} using sign-definite integral quantities that no solution of the limiting equation (\ref{third-order-limit}) with the boundary conditions (\ref{decreasing}) exists.

As a result, the geometric theory implies that the increasing shock (\ref{properties-h})
centered at $\theta = -\theta_0$ persists as a smooth solution $h(\theta)$ of the third-order equation
(\ref{third-order}) for any small $\epsilon > 0$, whereas the
decreasing shock centered at $\theta = \theta_0$ does not persist
in the third-order equation (\ref{third-order}) for any small $\epsilon > 0$.

This conclusion holds for any fixed $M > M_*$. It does not exclude, however, a
possibility of a complicated branching behavior in the solutions of the third-order
equation (\ref{third-order}) that can come from $M = \infty$ for small values of $\epsilon > 0$.
We shall consider the construction of solutions of the third-order equation (\ref{third-order})
numerically.

\section{Numerical analysis}

We shall construct numerical approximations of solutions of the third-order differential equation
(\ref{third-order}). The numerical approximations were generated using a custom-written turning-point algorithm and implemented in MATLAB. Solutions were found using Newton-Raphson iterations using Fourier spectral differentiation matrices with 256, 512, and 1024 Fourier modes. The mass--flux diagram was generated with parameter continuation of $Q$ or $M$, as decided by the algorithm. A convergent solution was defined numerically if the $(n+1)^{\rm th}$ iteration $h_{n+1}(\theta)$ satisfied
$$
\sup_{\theta \in [-\pi,\pi]}|h_{n+1}(\theta) - h_n(\theta)| \leq 5 \times 10^{-8}.
$$
Eventually non-convergent solutions were found and the parameter continuation failed at this `turning point'. To resolve the mass--flux diagram near the turning point, the following algorithm was implemented: first, a reference point was identified in $k$ steps behind the turning point (for our simulations, we generally chose $k=5$). Next, the convergence of the points making up a half-circle centered at the turning point was checked. The orientation of the half-circle was chosen to be facing away from the direction of the current parametrization (i.e. if we are increasing along the vertical axis then the lower half of the circle would be chosen). A vector was then drawn from the turning point to the convergent point with the largest distance from the reference point. Finally, the direction of new parametrization was chosen from the largest component of this vector. We found that this algorithm successfully navigated the loops in the mass-flux diagram.

\begin{figure}
\begin{center}
\includegraphics[height= 7 cm] {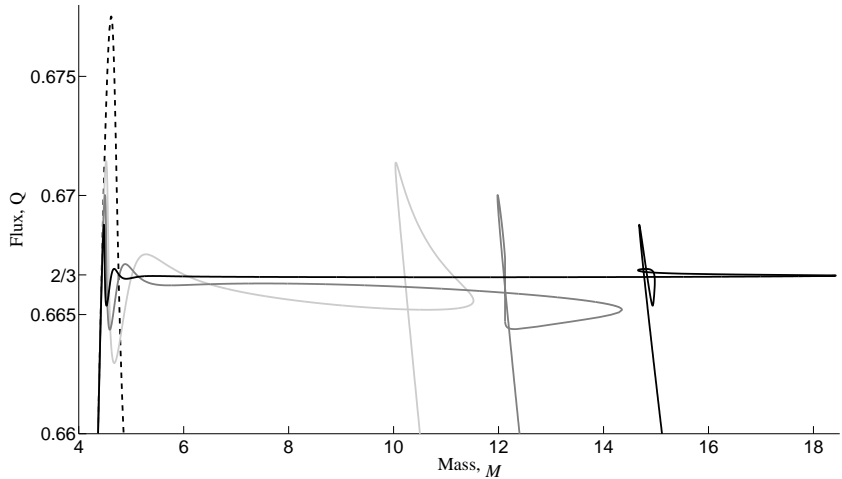}
\end{center}
\caption{The mass ($M$) versus the flux ($Q$) of the steady state
solutions of a thin film on the interior of a rotating cylinder for
various values of $\epsilon$: $0.005$ (dashed), $0.001$ (light gray),
$0.0005$ (dark gray), and $0.0001$ (black). }
\label{figure-1}
\end{figure}

Figure \ref{figure-1} shows the mass-flux diagram of stationary solutions for four values of $\epsilon$.
For $\epsilon = 0.005$ (dashed curve), we see no loops in the mass-flux diagram. For each fixed value of
mass $M$, there is exactly one value of the flux $Q$ for the stationary solution of (\ref{third-order}).
The first loop is formed for $\epsilon = 0.001$ (light gray). In an interval of values of $M$, three
stationary solutions coexist for three different values of $Q$. Note that this loop was discovered by
Benilov {\em et al.} \cite{Benilov3} (Figure 14).

Reducing $\epsilon$ further, we observe a formation and persistence of the second loop
in the mass-flux diagram for $\epsilon = 0.0005$ (dark gray) and $\epsilon = 0.0001$ (solid black).
Five solutions for different values of $Q$ coexist in an interval of values of $M$. The number of loops
keeps increasing as $\epsilon$ decreases to zero and their location is drifted to large values of $M$.
As $\epsilon \to 0$, the mass-flux diagram represent an increasing curve for $M \in (0,M_*)$ and
a constant level $Q = Q_*$ for $M > M_*$, where $Q_* = \frac{2}{3}$ and $M_* \approx 4.446$.
The limiting picture corresponds to the mass-flux diagram of the solutions of the limiting equation (\ref{algebraic-order})
that include the smooth solutions for $Q < Q_*$ and the shock solutions for $Q = Q_*$.

\begin{figure}
\begin{center}
\includegraphics[height= 7cm] {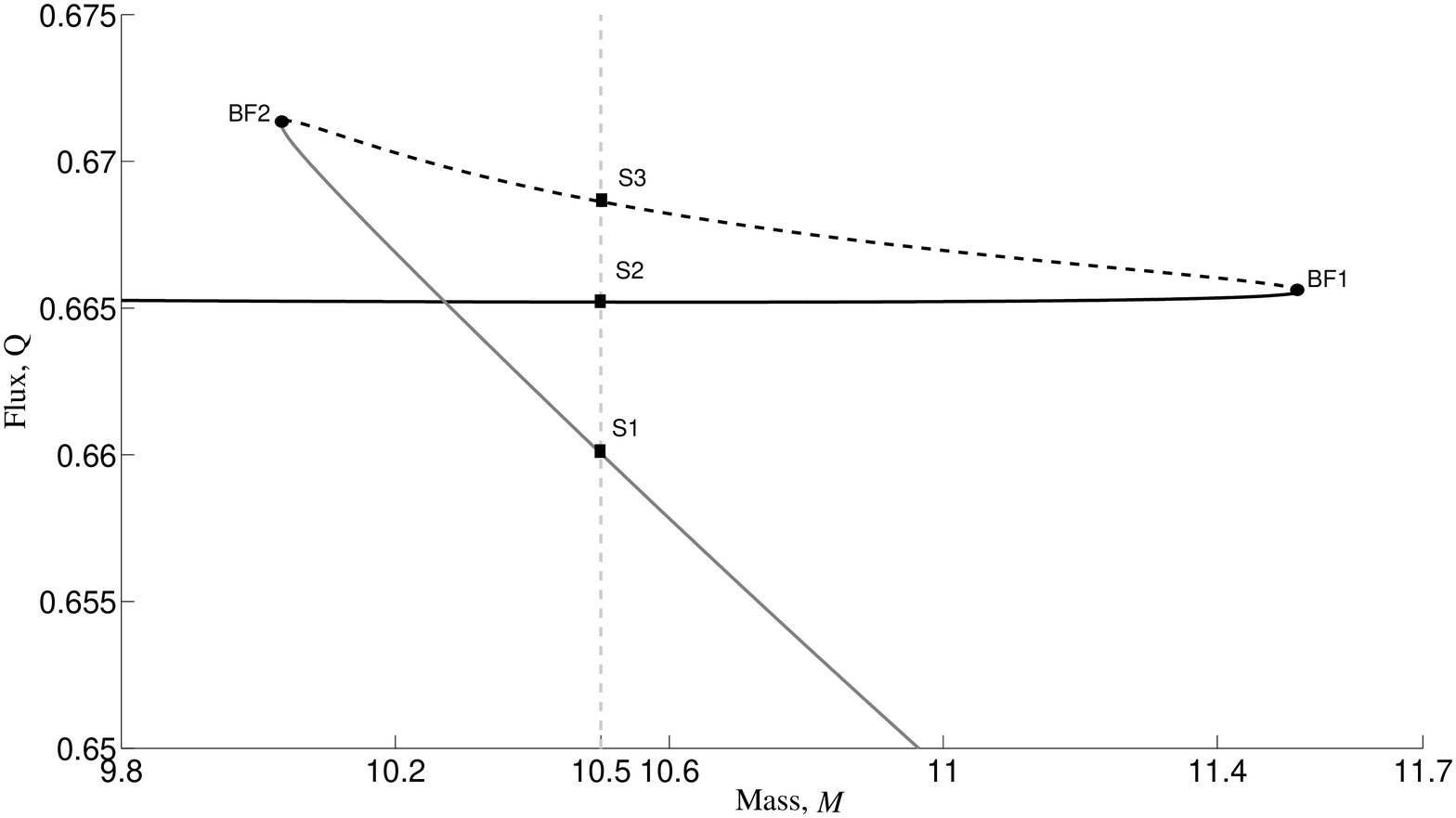} \\
\includegraphics[height= 7cm] {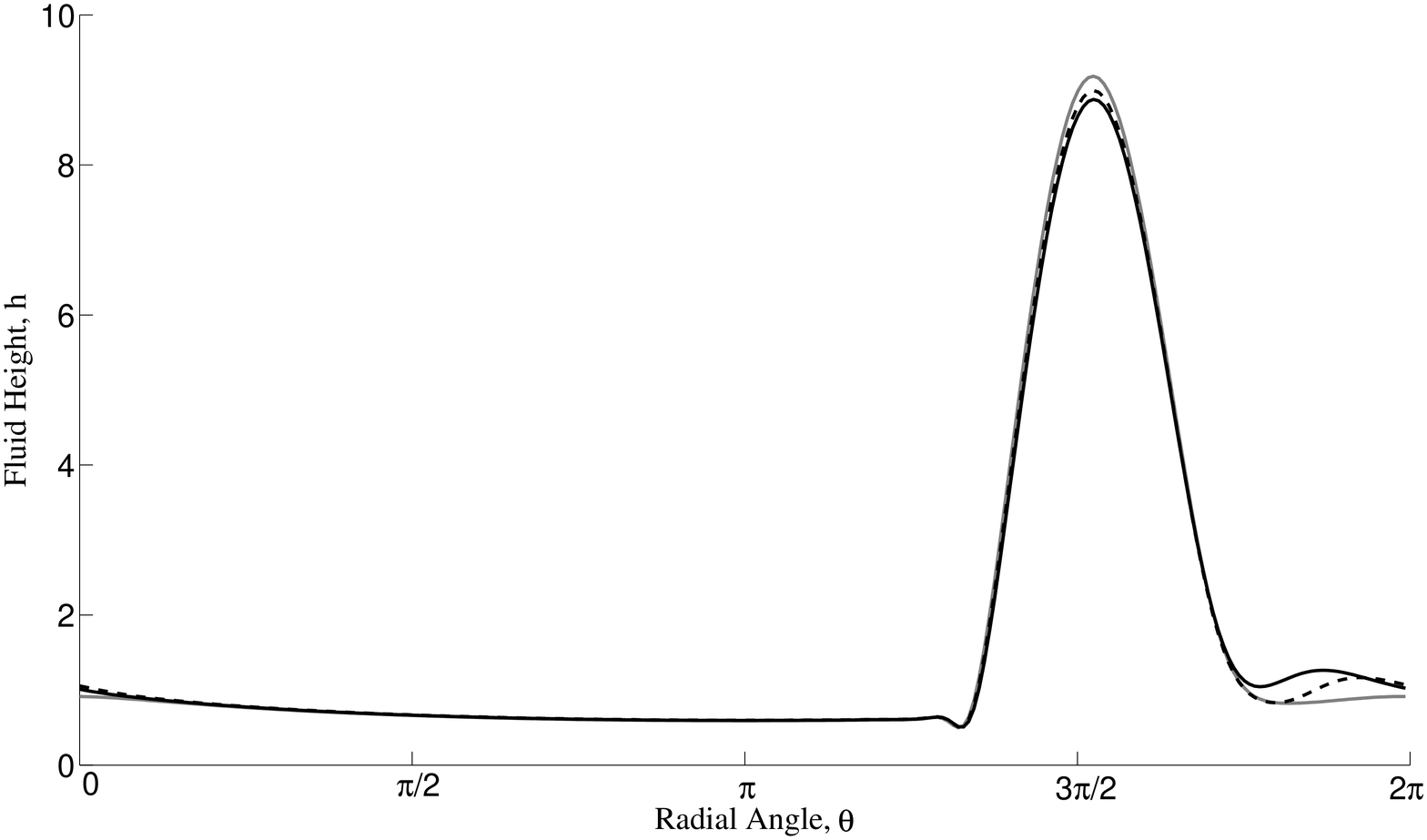}
\end{center}
\caption{(Top) A segment of the mass-flux diagram for $\epsilon =
0.001$. Three branches are indicated between two bifurcation points
(labeled BF1 and BF2). The three branches are shown by solid gray line (S1),
solid black line (S2), and dashed line (S3). The dashed gray line
shows the value of mass $M = 10.5$. (Bottom) Three steady state solutions with $M = 10.5$ and
$\epsilon = 0.001$: S1 (solid gray line, $Q = 0.6601$), S2 (solid black line, $Q = 0.6652$),
and S3 (dashed line, $Q = 0.6686$).}
\label{figure-2}
\end{figure}

We focus now on two particular examples of the mass-flux diagram with a single loop and a double loop.
Figure \ref{figure-2} (top) shows the mass-flux diagram for $\epsilon = 0.001$ with a single loop.
We can identify three solution branches (labeled as S1,S2, and S3) connected at two bifurcation
points (labeled as BF1 and BF2). The other point of intersections of solution branches S1 and S2 is not
a bifurcation point because the two solutions for the same value of $M$ and $Q$ remain distinguishable
into two different solutions. For $M = 10.5$, we compute the solution profiles and show them
on Figure \ref{figure-2} (bottom). Although similar in their shapes, the three steady
state solutions are clearly distinct. The peaks of the solutions are located for $\theta > \pi$,
or equivalently for $\theta < 0$, thanks to the $2\pi$-periodicity of the solutions.
They correspond to the increasing shock solution as $\epsilon \to 0$ located at $\theta = -\theta_0 < 0$.
Oscillations, which are visible on both sides of the shock are attributed to complex eigenvalues
of the linearized equation (\ref{linearized}) after the scaling transformation
(\ref{scaling-transf}) and the limit $\epsilon \to 0$.

Because of multiple steady-state solutions with the same physical parameter of the mass $M$,
we anticipate that they may have different stability properties. Therefore, we examine eigenvalues
of the linearized equation
\begin{equation}
\label{lin-third-order}
\lambda f + \partial_{\theta} \left[ f - h^2 \cos(\theta) f
+ \epsilon h^2 (\partial_{\theta} h + \partial_{\theta}^3 h) f +
\frac{1}{3} \epsilon h^3 (\partial_{\theta} f + \partial_{\theta}^3 f) \right] = 0,
\end{equation}
where $h(\theta)$ is a $2\pi$-periodic steady-state solution and
$f(\theta)$ is a $2\pi$-periodic perturbation to the steady state with the growth
rate $\lambda \in \mathbb{C}$. If ${\rm Re}(\lambda) > 0$, the perturbation
$f(\theta) e^{\lambda t}$ grows on the background of $h(\theta)$ and induces instability of
the steady-state solution. Expressing $h(\theta)$ from the third-order equation (\ref{third-order}),
we can rewrite the spectral problem (\ref{lin-third-order}) in the equivalent form,
\begin{equation}
\label{lin-third-order-new}
L f = \lambda f, \quad L = \frac{\partial}{\partial \theta} \left[ \frac{2h-3Q}{h} -
\frac{1}{3} \epsilon h^3 \left( \frac{\partial}{\partial \theta}  +
\frac{\partial^3}{\partial \theta^3} \right) \right].
\end{equation}

Note that there is always a zero eigenvalue in the spectral problem
(\ref{lin-third-order-new}). Indeed, $f_0 = \frac{\partial h}{\partial Q}$ is an eigenfunction
for $\lambda = 0$ because $Q$ is a free parameter of the solution $h$.
We will now show that the zero eigenvalue is simple. First, since the Jacobian operator 
in the iteration algorithm was found to be invertible, the operator $L_0$ is invertible, where
$$
L_0 = \frac{2h-3Q}{h} - \frac{1}{3} \epsilon h^3 \left( \frac{\partial}{\partial \theta}  +
\frac{\partial^3}{\partial \theta^3} \right).
$$
Since $L = \partial_{\theta} L_0$ and $\partial_{\theta}$ has a one-dimensional kernel, 
the operator $L$ has at most one eigenvector in the kernel, which is 
$f_0 = \frac{\partial h}{\partial Q}$. 

Next, we consider the adjoint spectral problem,
\begin{equation}
\label{lin-third-order-adjoint}
L^* g = \lambda g, \quad L^* = \left[ \frac{3Q - 2h}{h}  -
\frac{1}{3} \epsilon \left( \frac{\partial}{\partial \theta}  +
\frac{\partial^3}{\partial \theta^3} \right) h^3 \right] \frac{\partial}{\partial \theta}.
\end{equation}
It is clear that $g_0 = 1$ is the adjoint eigenfunction for $\lambda = 0$ and that
$$
\langle g_0,f_0 \rangle = \frac{1}{2\pi} \int_{-\pi}^{\pi} \frac{\partial h}{\partial Q} d \theta =
\frac{dM}{dQ}
$$
is nonzero at all values of $Q$ but the bifurcation points BF1 and BF2. By Fredholm' theory 
for isolated eigenvalue, this fact implies that $\lambda = 0$ is a simple eigenvalue for 
all values of $Q$ but the bifurcation points BF1 and BF2.

We use a numerical method based on building a matrix representation of the
differential operator $L$ acting on $f$. To do so, we discretize the space and approximate the derivatives using the Fourier spectral method. The eigenvalue problem is then solved using the MATLAB function
{\it eig}.

\begin{figure}
\begin{center}
\includegraphics[height= 7 cm] {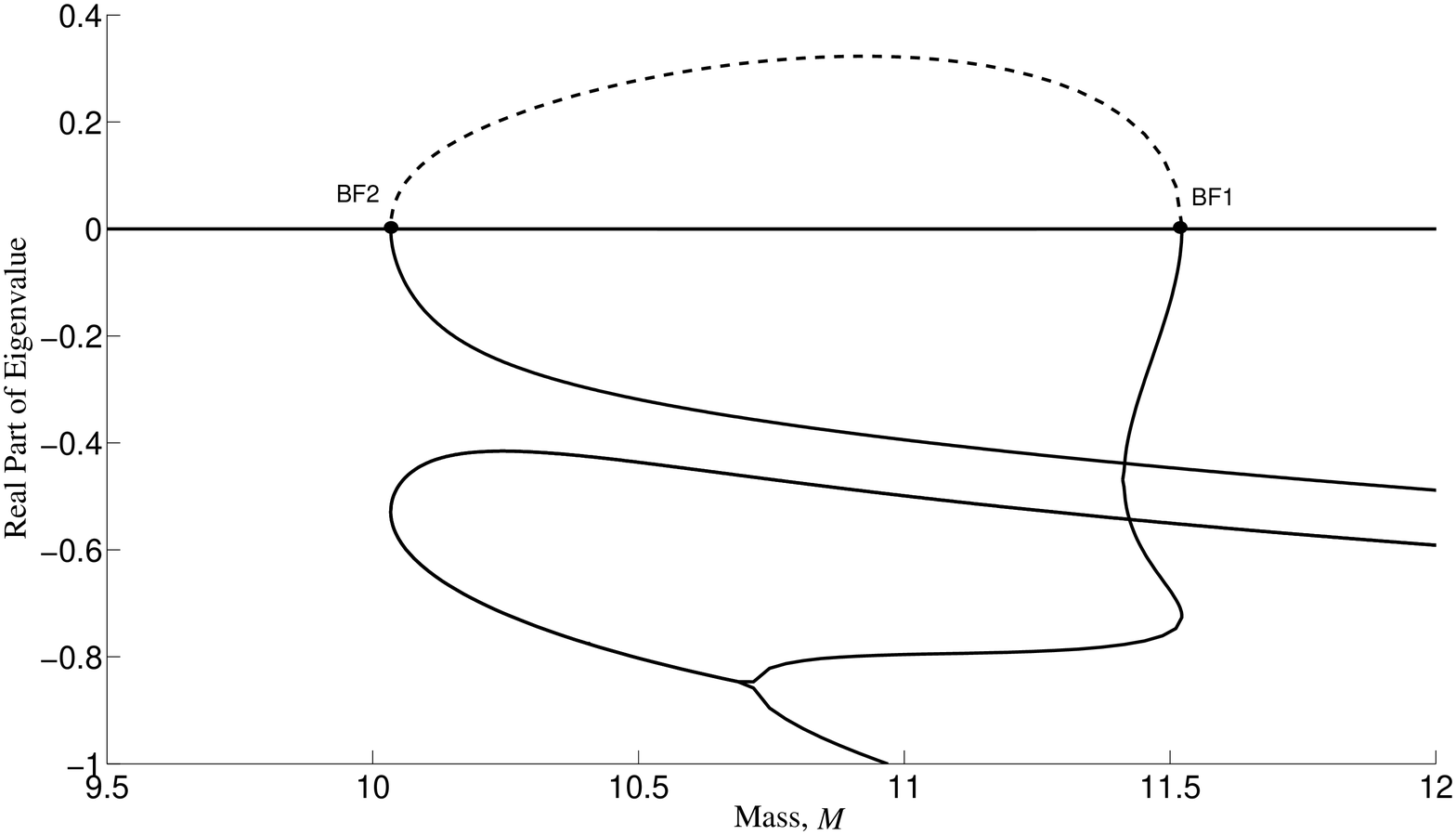}
\end{center}
\caption{The real part of the smallest eigenvalues $\lambda$ of the spectral problem (\ref{lin-third-order})
for $\epsilon = 0.001$ and $M = 10.5$.}
\label{figure-4}
\end{figure}

The real part of the smallest eigenvalues $\lambda$ of the spectral problem (\ref{lin-third-order-new})
is shown on Figure \ref{figure-4} for $\epsilon = 0.001$ and $M = 10.5$. All other eigenvalues
have larger negative real parts. Between the two bifurcation points
BF1 and BF2, one eigenvalue crosses zero and becomes unstable,
revealing a saddle-node bifurcation at points BF1 and BF2. Branch S3
(between BF1 and BF2) is unstable with exactly one real positive eigenvalue $\lambda$.
Other two branches S1 and S2 are stable with all but one zero eigenvalue having
negative real parts. It follows from Figure \ref{figure-2} (bottom) that the
profile of the steady-state solution at branch S3 is
squeezed between those of the solutions at branches S1 and S2.
Small perturbations of the middle tail of
the unstable solution is expected to grow towards the upper
or lower tails of the other two stable solutions.
Particular values of $\lambda$ for branches S1, S2, and S3 are
given in Table 1.

\begin{center}
\begin{tabular}{|l|l|l|}
\hline
solution branch & real $\lambda$ & complex $\lambda$  \\
\hline S1, $Q = 0.6601$ & $-0.32$ &  $-0.44 \pm i 0.79$, $-0.72 \pm i 1.50$ \\
  \hline
  S2, $Q = 0.6652$ & $-1.34$ & $-0.75 \pm i 0.44$, $-1.31 \pm i 0.97$ \\
   \hline
  S3, $Q = 0.6686$ & $0.28$ & $-0.80 \pm i 0.11$, $-1.18 \pm i 1.07$ \\
  \hline
\end{tabular}

{\bf Table 1:} Smallest nonzero eigenvalues of the spectral problem (\ref{lin-third-order})
for $\epsilon = 0.001$ and $M = 10.5$.
\end{center}

\begin{figure}
\begin{center}
\includegraphics[height= 7 cm] {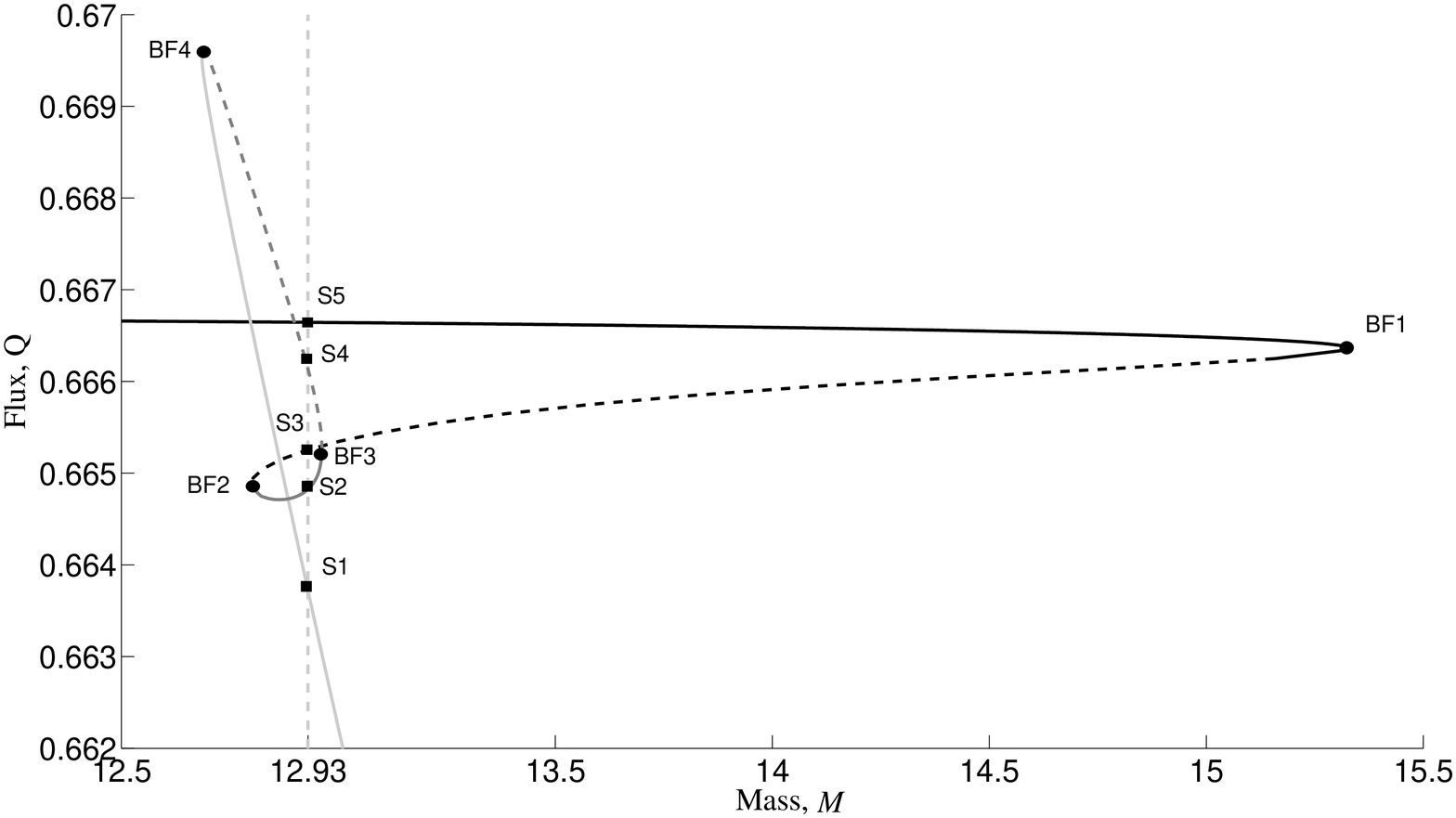} \\
\includegraphics[height= 7 cm] {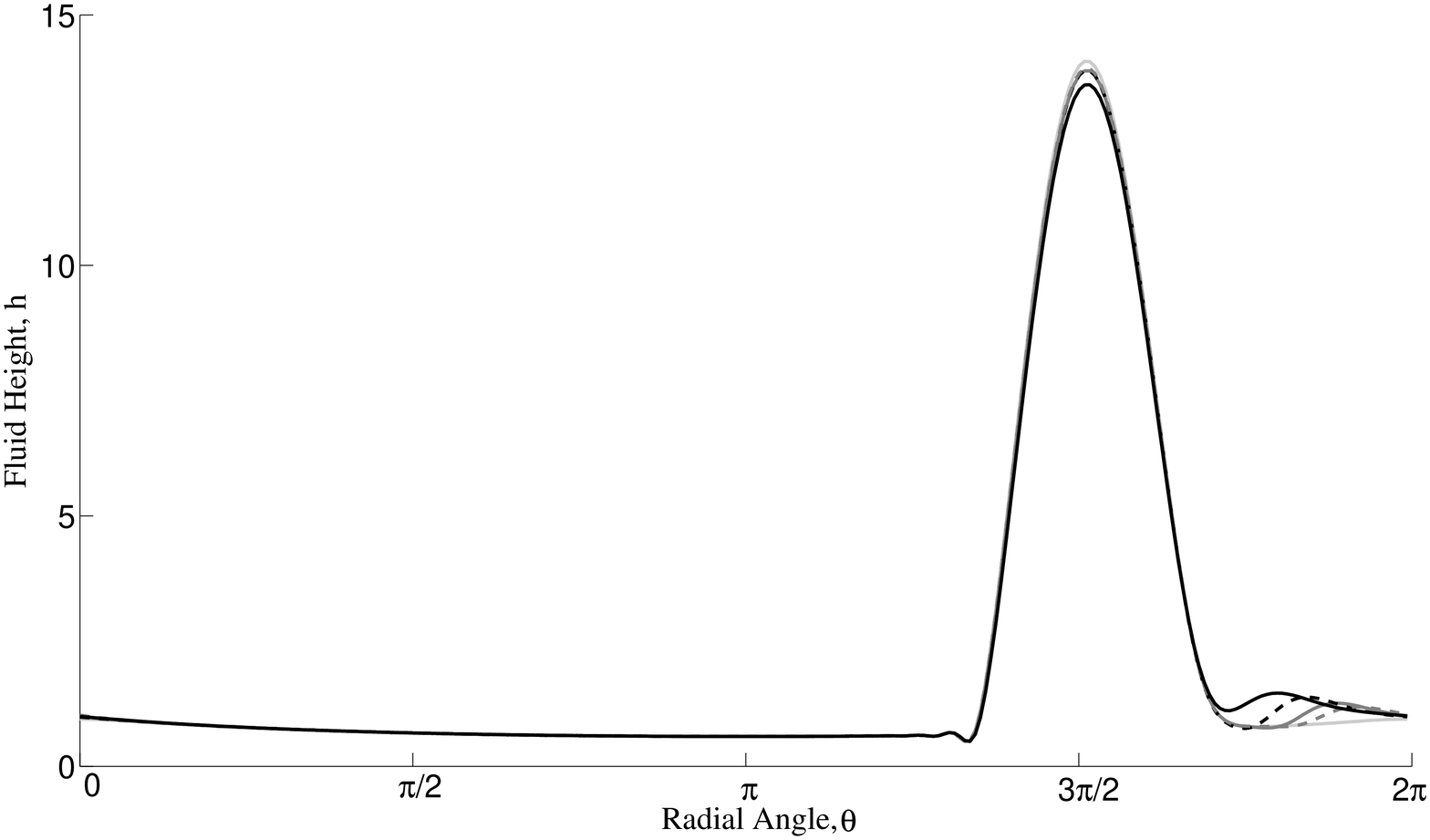}
\end{center}
\caption{(Top) A segment of the mass-flux diagram for $\epsilon =
0.00039$. Five branches are indicated between four bifurcation points
(labeled BF1, BF2, BF3, and BF4 and shown by circles). The  five branches are shown by solid light gray line (S1),
solid dark gray line (S2), dashed line (S3), dashed gray line (S4), and solid black line (S5). The dashed gray line
shows the value of mass $M = 12.93$. (Bottom) Five steady state solutions with $M = 12.93$ and
$\epsilon = 0.00039$: S1 (solid light gray line, $Q = 0.6638$), S2 (solid dark gray line, $Q = 0.6648$), S3
(dashed line, $Q = 0.6653$), S4 (dashed gray line, $Q = 0.6661$), S5
(solid black line, $Q = 0.6666$).}
\label{figure-5}
\end{figure}

\begin{figure}
\begin{center}
\includegraphics[height= 7 cm] {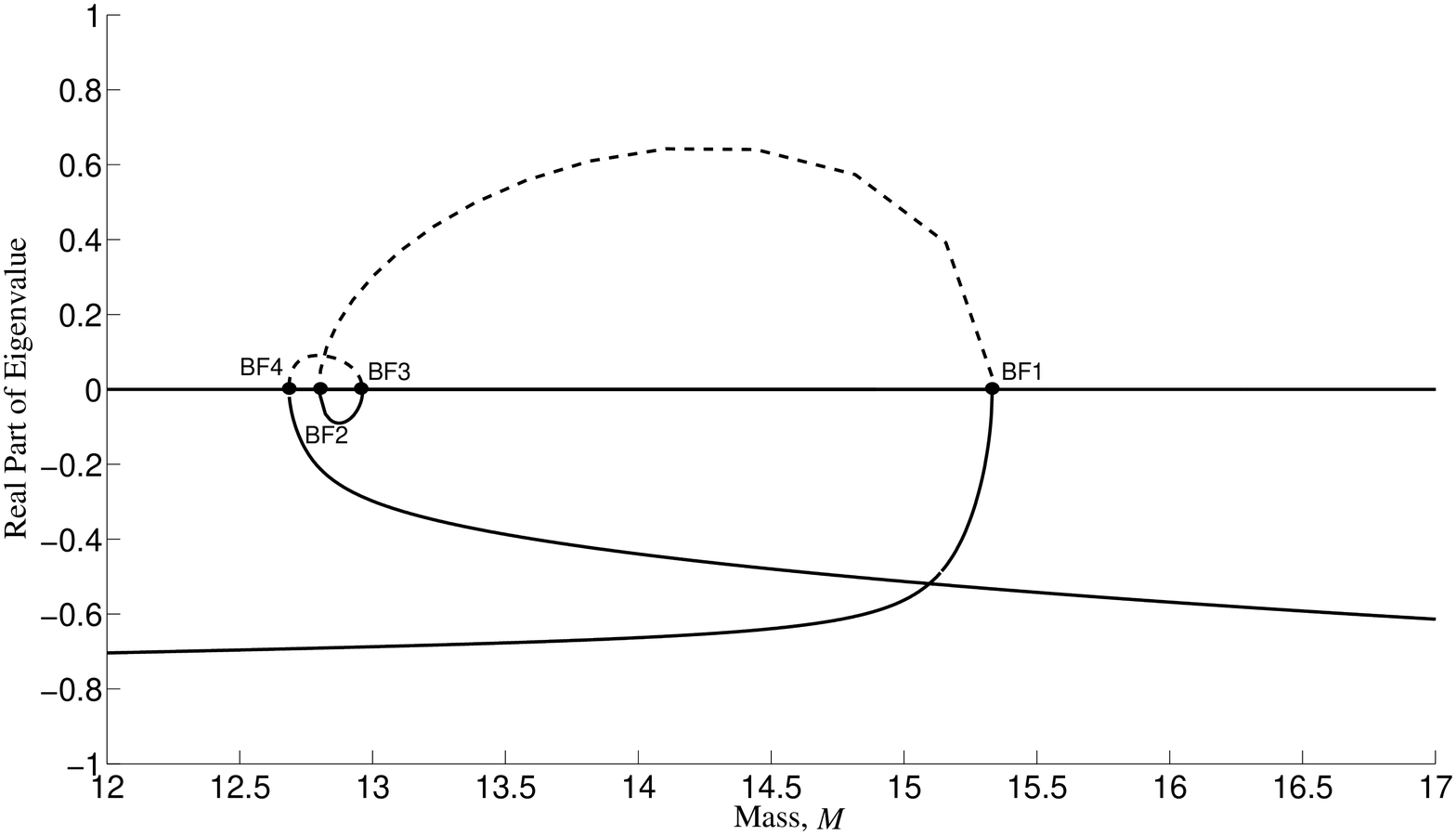}
\end{center}
\caption{The real part of the smallest eigenvalues $\lambda$ of the spectral problem (\ref{lin-third-order})
for $\epsilon = 0.00039$ and $M = 12.93$.}
\label{figure-7}
\end{figure}

Figure \ref{figure-5} (top) shows the mass-flux diagram for $\epsilon = 0.00039$ with two loops.
We can identify five solution branches (labeled as S1,S2,S3,S4, and S5) connected at four bifurcation
points (labeled as BF1, BF2, BF3, and BF4). For $M = 12.93$, we compute the solution profiles and show them
on Figure \ref{figure-5} (bottom). Although similar in their shapes, the five steady
state solutions are clearly distinct. Properties of these solutions resemble those on Figure \ref{figure-2}.
In particular, multiple steady states exist for a fixed mass $M$ and $\epsilon$ and can be
identified by their flux values $Q$. The five steady-state solutions are
almost identical with the most visible deviation in their tails and peak heights.

Comparison between Figures \ref{figure-2} and \ref{figure-5} shows that how as $\epsilon$
decreases, more loops on the mass-flux diagram form and the number of steady states increases.

Although the five steady-state solutions exist mathematically, whether or not they could exist
physically depends on their stability. The real part of the smallest eigenvalues $\lambda$
of the spectral problem (\ref{lin-third-order}) is shown on Figure \ref{figure-7}
for $\epsilon = 0.00039$ and $M = 12.93$. All other eigenvalues
have larger negative real parts. Between the bifurcation points
BF1 and BF2, BF2 and BF3, BF3 and BF4, one eigenvalue crosses zero,
revealing four saddle-node bifurcation at these points. Branch S3
(between BF1 and BF2) is unstable with exactly one real positive eigenvalue $\lambda$.
Branch S4 between BF3 and BF4 is also unstable with exactly one real positive eigenvalue.
Other three branches S1, S2, and S5 are stable with all but one zero eigenvalue having
negative real parts. Again, we point readers to
Figure \ref{figure-5} (bottom) that shows how tails of unstable solutions S3 and S4 are aliased by
the tails of stable solutions S1, S2, and S5.
Particular values of $\lambda$ for branches S1, S2, S3, S4, and S5 are
given in Table 2.

\begin{center}
\begin{tabular}{|l|l|l|}
  \hline
   branch & real $\lambda$ & complex $\lambda$  \\
  \hline
  S1, $Q = 0.6638$ & $-0.28$, $-3.76$ &  $-0.42 \pm i 0.70$, $-0.65 \pm i 1.41$, $-0.90 \pm i 2.25$ \\
  \hline
  S2, $Q = 0.6648$ & $-0.07$, $-0.65$ & $-1.06 \pm i 0.82$, $-1.33 \pm i 1.82$, $-1.62 \pm i 2.86$ \\
   \hline
  S3, $Q = 0.6653$ & $0.30$, $-0.81$ &  $-1.17 \pm i 0.55$, $-1.50 \pm i 1.56$, $-1.84 \pm i 2.68$ \\
    \hline
  S4, $Q = 0.6661$ & $0.05$, $-0.69$  &  $-1.03 \pm i 0.96$, $-1.25 \pm i 1.94$, $-1.52 \pm i 2.94$ \\
      \hline
  S5, $Q = 0.6666$ & $-0.69$, $-2.57$ &  $-1.33 \pm i 0.46$, $-1.54 \pm i 1.46$, $-2.31 \pm i 1.32$ \\
  \hline
\end{tabular}

{\bf Table 2:} Smallest nonzero eigenvalues of the spectral problem (\ref{lin-third-order})
for $\epsilon = 0.00039$ and $M = 12.93$.
\end{center}

\section{Conclusion}

We have thus explored the behavior of a thin liquid film in a rotating cylinder
accounting for surface tension and gravity. In particular, we showed that
regularized increasing shock solutions persist under small surface tension.
These shock solutions were then visualized for a wide range of parameter values
by using numerical discretizations on an uniform grid. We have identified a number
of solution branches on the mass-flux diagram and have shown numerically
that the number of branches increases and the location of shocks move to infinity
as the surface tension decreases to zero.

We conclude by listing a number of open questions for further studies.
First, it is suggested by the numerical computations that the number of
solution branches goes to infinity as $\epsilon \to 0$ but computations
become difficult and unreachable for $\epsilon < 10^{-5}$. Second, the
steady state are expected to persist with respect to small inclinations
of the cylinder \cite{Russian} but we do not include inclined cylinders in this work.
Finally, numerical discretizations on the adaptive (variable) grid can be developed
further to resolve better the regularized shock solutions near the shock location.

\end{document}